\documentclass{llncs}

\usepackage{graphicx}
\usepackage{hyperref}
\usepackage{orcidlink}

\begin{document}

\pagestyle{plain}
\title{Large Language Models as Supervised Extraction Assistants: Lowering the Barrier to Documentation Standard Adoption in Agent-Based Modelling}

\author{Peer-Olaf Siebers\orcidlink{0000-0002-0603-5904}\inst{1}
\and Christopher Frantz\orcidlink{0000-0002-6105-8738}\inst{2}}

\institute{University of Nottingham, Nottingham, UK\\
\email{peer-olaf.siebers@nottingham.ac.uk}
\and
Norwegian University of Science and Technology (NTNU), Gj\o{}vik, Norway\\
\email{christopher.frantz@ntnu.no}\\[2mm]
\includegraphics[width=3cm]{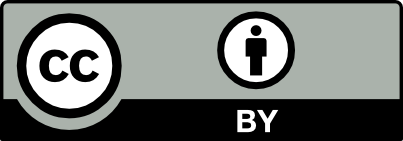}}
\maketitle

\begin{abstract}
Agent-Based Modelling (ABM) relies on clear documentation to ensure credibility and transparency. Although standards exist for documenting models (e.g. ODD), processes (e.g. TRACE, EABSS), and data use (e.g. RAT-RS), their adoption remains limited due to the effort required to produce documentation that is often treated as supplementary. This paper explores the use of Large Language Models (LLMs) to facilitate and partially automate such processes. We conduct a feasibility study focusing on the underused Rigour and Transparency Reporting Standard (RAT-RS), using four LLMs to extract reports from a published ABM paper. We assess consistency and performance across question types, finding that LLMs generate coherent outputs and perform more reliably on descriptive than on explanatory or evaluative tasks. While LLMs can improve reporting quality and consistency, they also exhibit notable limitations. We identify practical heuristics for when LLM-assisted documentation is reliable and when human oversight is needed and call for systematic community-level exploration to enhance rigour and adoption in ABM reporting.
\end{abstract}


\section{The Problem: Many Standards, Limited Adoption}
\label{sec:problem}

The credibility of social simulation as a method relies on transparent reporting of methodological choices, including model specification and, given its ability to integrate rich sets of data sources, the use of data employed as part of the modelling process. When sharing findings from Agent-Based Modelling (ABM) studies, researchers need to document what data were used, why they were chosen, and how they fit into the modelling process. This transparency is not just best practice; it is fundamental for reproducibility. In reality, however, this standard is often not met. With its long-standing presence in the community, the ODD (Overview, Design concepts, and Details) protocol by Grimm et al.\ \cite{grimm2020}, as de facto standard for ABM documentation, is still not consistently provided alongside models. This challenge is ongoing, despite increasing efforts to mandate the provision of model specifications for journal submission. Another emerging standard, the RAT-RS (Rigour And Transparency Reporting Standard) by Achter et al.\ \cite{achter2022} was created to serve as a complementary protocol to offer transparency about data use. It offers a structured set of questions covering specification, calibration, and validation, and it can accommodate quantitative, qualitative, and mixed-methods data sources. For both standards, anecdotal evidence and surveys consistently highlight effort as the main barrier, a barrier likely further exacerbated by evolved specifications (e.g. the ODD+D protocol by M\"{u}ller et al.\ \cite{muller2013}) that add further demands to the documentation.

The challenge to establish a consistent adoption of such community standards resembles a classic Tragedy of the Commons: the collective benefits of widespread adoption are manifold, including rigour in presenting individual models, affording more rapid assessment of credibility by affording data traceability, leading to more likely adoption of models for further adaptation or application in real-world settings. Moving beyond individual models, a consistent adoption would further strengthen the credibility of social simulation as a method across a broader range of disciplines and enable novel avenues of large-scale comparative research. At the same time, individual cost (in terms of time) and the limited immediate benefit as prima facie supplementary information (rather than core content) deter engagement. 

Reflecting on these observations, a key question is whether we can develop novel strategies to encourage community adoption by reducing the effort involved in doing so. We claim that one such opportunity avails itself with the increasing availability of Large Language Models (LLMs), alongside initial evidence that they are candidates to support efforts previously exclusively performed by humans. 

To this end, this paper explores based on a feasibility study to which extent LLMs can support supervised data extraction and augmentation from publications for model documentation (with particular focus on data use). We articulate our position in Section~\ref{sec:position}. In Section~\ref{sec:case} we introduce our exemplary case alongside a general methodology that can be adapted to different purposes. Sections~\ref{sec:cross} and~\ref{sec:intern} explore how manual extraction compares to automated extraction to humans and across different models, alongside robustness tests to ensure reliable reproducibility. Augmenting these primarily quantitative insights, in Section~\ref{sec:exam}, we explore how the quality of data extraction varies by question/item type, providing a more general insight into circumstances under which the use of LLMs is appropriate (and variably less so). The associated assumptions and caveats are provided in Section~\ref{sec:caveats}. In Section~\ref{sec:discussion} we turn to a concluding discussion that paves a path forward, both in terms of technical development, as well as opportunities and implications for the adoption of LLM more broadly, before translating our observations into a call for action addressed at the community in the concluding Section~\ref{sec:action}. 

\section{The Position: Supervised Extraction at Scale}
\label{sec:position}

We believe LLMs, when used under human supervision, can make standard and protocol adoption practical on a larger scale. We refer to this approach as supervised extraction. The central idea is that the LLM is involved in an assistive facility, performing initial data collection work by scanning the paper, identifying relevant sections, organising information according to given protocol templates, and noting where evidence is missing or unclear. The human researcher then validates, corrects, and approves the output, hence staying in control of the overall process. This shifts the researcher's role from producer to constructive critic, or reviewer. While reducing the cognitive load associated with this task, it bears the potential to resolve the dissonance that can be involved when developing the documentation on the one hand, while ensuring quality based on self-assessment. The supervised extraction approach is not entirely new. It follows a common model seen in related fields: systematic review automation (e.g.\ Schroeder et
al.\ \cite{schroeder2025}), clinical data extraction (e.g.\ Abdellaoui et al.\ \cite{abdellaoui2025}), evidence synthesis (e.g.\ Gartlehner et al.\ \cite{gartlehner2024}), as well as initial adoption for the purpose of post-hoc extraction of model documentation (e.g.\ Khatami and Frantz \cite{khatami2025}) have moved towards human-in-the-loop LLM workflows, showing improvements in both speed and consistency. An application that we introduce here is to develop a complementary approach for ABM data use reporting, with specific focus on the RAT-RS as underlying standard. We choose this particular example, since it is underexplored, yet strongly relevant to establish rigorous documentation, and due to its comparatively fine-granular nature based on 39 structured questions, making it an ideal candidate to explore the boundaries of such LLM-assisted information extraction. 

As motivated above, the employed analogy in the systematic engagement of LLMs is one of a junior researcher. A junior researcher tasked with documenting data use in a published paper might draft a report that is generally accurate but not without errors. You would not publish that draft without checking it, but you would not start from scratch either. The junior researcher has done the hard work of reading the paper thoroughly, identifying every mention of data sources, and trying to map that information onto a structured template. Your task is to verify, not unearth. LLMs, when used with a well-crafted prompt, can fulfil, or at least support, this role. This approach has practical implications: the barrier to RAT-RS adoption is no longer just effort. With an LLM-assisted workflow, the researcher's time commitment might be substantively reduced, while potentially revealing potential gaps or blind spots in documentation, as well as help establish reporting consistency based on standardised prompts. In a field where time pressures can lead to shortcuts in (or absence of) documentation, the proposed approach can lead to substantive opportunities, both in terms of quality and uptake, if performed in a responsible and systematic manner.

\section{Case and Methodology}
\label{sec:case}

\subsection{Supporting Evidence: A Feasibility Study}
\label{sec:feasibility}

The purpose of this work is not to advance a statistically generalisable claim or assessment of the proposed approach. Rather, it displays a controlled feasibility test with well-understood parameters, establishing the principal utility of LLMs to support documentation tasks. This enables a depth-first strategy to devise a methodological approach that (a) showcases (or refutes) the potential of LLM-assisted extraction methods, and (b) introduces mechanisms to establish the reliability and robustness of this approach.

We specifically chose RAT-RS for this evaluation (an aspect briefly motivated above), since it is representative for underused standards whose broader adoption has the potential to substantively increase rigour in data use reporting (a central feature of many ABM studies), with adoption challenged due to its rather granular and detailed nature. At the same time, its question-centred approach (as opposed to more open-ended categories) lends itself well for direct interaction with large language models, allowing us to not only identify the potential of this specific standard, but to draw broader insights related to different question categories, an aspect equally relevant if intending to semi-automate extraction of other protocols or standards. 

To test the effectiveness of supervised extraction, we base the study on Siebers and Aickelin's \cite{siebers2011} published paper 'A First Approach on Modelling Staff Proactiveness in Retail Simulation Models', a worked example included with the original RAT-RS publication. This paper was selected, since a manually completed RAT-RS report already exists, providing a human-authored dataset against which LLM responses could be compared. Conducted in March 2026, the feasibility study put four state-of-the-art LLMs to the test: Claude Sonnet 4.6, DeepSeek-V3, ChatGPT-5.1, and Gemini Flash 3, using their internal reasoning/thinking mode. Each was given the same modular extraction prompt and the same publication, generating RAT-RS reports through standard chat interfaces without fine-tuning or API-level configuration. The output analysis proceeded along two complementary dimensions: a cross human/LLM and LLM/LLM semantic analysis examining the degree of agreement across human and LLM responses and an intra-model semantic analysis assessing whether an individual model produces stable outputs across repeated runs of the same prompt.

\subsection{The RAT-RS}
\label{sec:ratrs}

\begin{figure}[htbp]
  \centering
  \includegraphics[width=0.8\linewidth]{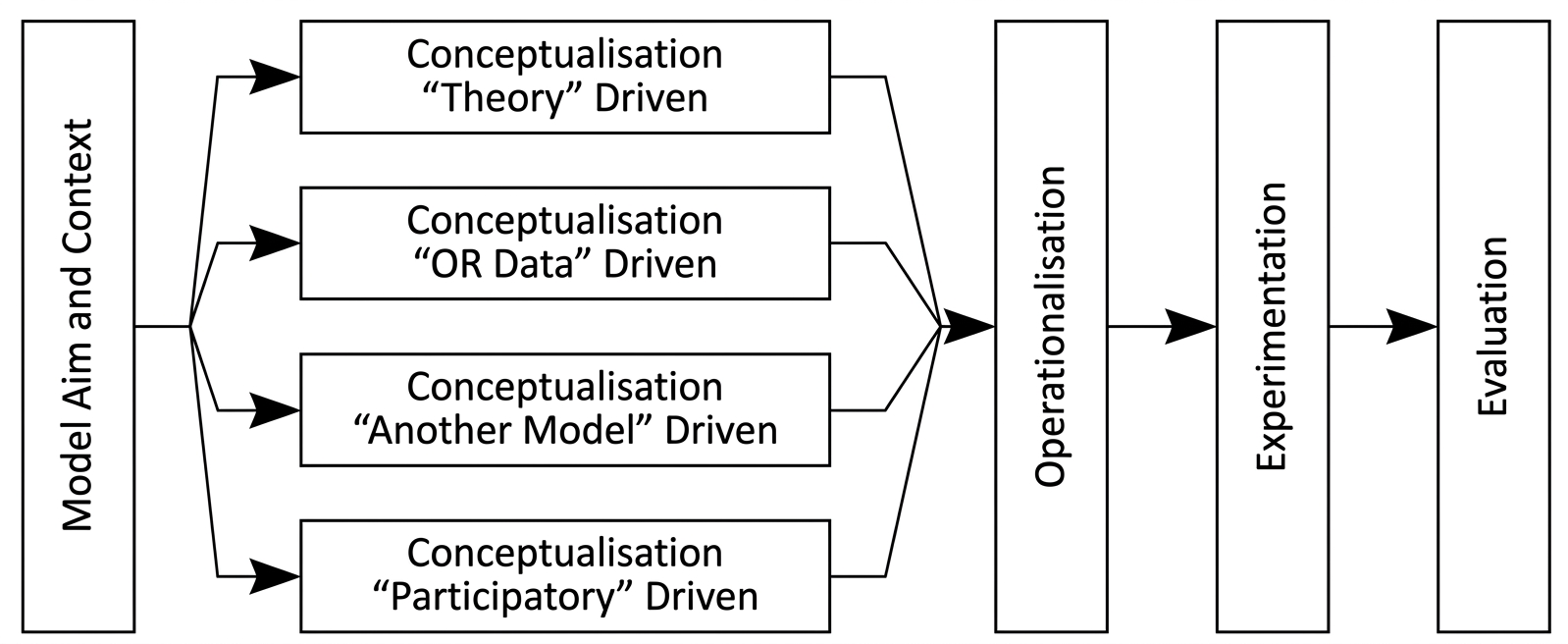}
  \caption{RAT-RS QS toolbox structure and workflow.}
  \label{fig:ratrs}
\end{figure}

The RAT-RS is a reporting standard published in Achter et al.\ \cite{achter2022}. It is intended to improve documentation of data use in ABM. It supports reporting across several development stages of an ABM (specification, calibration, and validation) and is compatible with a variety of data types, including statistical, qualitative, ethnographic and experimental data, consistent with mixed methods research. The RAT-RS is organised as a Question Suite (QS) toolbox (Figure~\ref{fig:ratrs}). Each QS focuses on a distinct aspect of data use. There are multiple RAT-RS "flavours" with distinct Conceptualisation QSs.

The first step in applying the RAT-RS is to identify the main driver of model development. The RAT-RS supports four approaches: theory-driven models (focusing on pre-existing theories), OR-data-driven models (focusing on key mechanisms), another-model-driven models (focusing on pre-existing models), and participatory-driven models (focusing on participatory design processes), as visualised below.

For the remainder of this text, whenever we mention RAT-RS, we refer to the OR Data flavour version (emphasis on key model mechanisms), as this reflects the case of the paper we are using in this feasibility study. The original RAT-RS consists of 39 questions distributed into five different QS representing the entire ABM cycle. For the experimentation with LLM we extended the original RAT-RS by adding a sixth QS with five additional questions that have meta-evaluative function in assessing the quality of data reporting in the publication and providing advice for improvement or to reveal gaps in reporting - an aspect that provides tangible instructive guidance sponsored by the interactive nature of LLM-based augmentation. This additional interactive, and potentially iterative, mode moves beyond the post hoc documentation questionnaire of the original standard. The extended RAT-RS hence consists of 44 questions. The full set of questions is available for download from GitHub\footnote{\url{https://github.com/PeerOlafSiebers/ssc2026}}.

\subsection{Question Type Analysis}
\label{sec:qtypes}

The 44 RAT-RS questions vary considerably in the type of reasoning they require, a distinction that shaped the prompt design and underpins the consistency analyses that follow. We classified questions using a stem-based approach that identifies lexical and syntactic cues in the question stem and distinguished five types: (1) Factual: Requests an objectively verifiable datum with a unique correct answer that can be confirmed externally without reasoning or judgement. (2) Descriptive: Asks the respondent to characterise or enumerate model features using "what" or "which" interrogatives, with answers locatable directly within the paper. (3) Explanatory: Requires causal or rationale-based reasoning to justify a decision or choice, typically cued by "why" or "explain". (4) Binary: Elicits a dichotomous yes/no response, often followed by conditional elaboration depending on that answer. (5) Evaluative: Calls for an external meta-level appraisal or judgement of the paper as a whole, rather than reporting or reasoning about its content directly. Using the above approach, we have classified the RAT-RS questions as follows: Factual: Q1.1; Descriptive: Q1.2, Q1.3, Q1.4, Q1.5, Q1.7, Q2.1, Q2.3, Q2.4, Q2.7, Q3.1, Q3.2, Q3.4, Q3.9, Q3.12, Q4.1, Q4.2, Q4.3, Q4.4, Q4.5, Q4.6, Q4.8, Q5.1, Q5.3; Explanatory: Q1.6, Q1.8, Q2.2, Q2.5, Q2.6, Q3.3, Q3.5, Q3.8, Q3.10, Q3.11, Q3.13, Q4.7, Q5.2; Binary: Q3.6, Q3.7; Evaluative: Q6.1, Q6.2, Q6.3, Q6.4, Q6.5.

\subsection{Prompt Design and Architecture}
\label{sec:prompt}

Building the extraction prompt was an iterative process, informed throughout by the current state-of-the-art in prompt engineering research. The prompt incorporates the following evidence-based design principles: modular architecture \cite{khot2022}; markdown formatting \cite{he2024}; task-specific persona \cite{white2023}; sequential step decomposition implementing \textit{chain-of-thought} prompting \cite{wei2022}; explicit grounding rules to suppress hallucination \cite{huang2025}; evidential basis signalling using a four-category epistemic taxonomy \cite{xiong2024}; self-verification via a structured meta-summary \cite{weng2023}; a constrained structured output format with a 100-word response limit \cite{vatsal2024}; in-prompt terminology definitions to reduce \textit{interpretive drift} \cite{vatsal2024,ziems2024}; a worked anonymised example to anchor output format through \textit{few-shot in-context learning} \cite{brown2020,min2022}; and strategic use of markdown \cite{he2024}.

The final prompt comprising five components: (1) Role: Establishes a task-specific expert persona, priming the model's domain focus and output register before any instruction is issued. (2) Instructions: Provide a structured eight-step protocol governing how the model reads the paper, classifies and extracts information, signals evidential certainty, handles conditional dependencies, and reflects critically on patterns of omission across the full question set. (3) Output Format: Specifies a four-column CSV table with strict 100-word response limits, response rules keyed to each evidential signal, and an anonymised worked example anchoring the expected structure. (4) Template: Presents the 44 RAT-RS questions across six thematic QSs, which the LLM completes using all preceding rules. (5) Appendix: Supplies a terminology table defining eight domain-specific terms, reducing interpretive drift across the full question set. The full RAT-RS extraction prompt is available for download from GitHub\footnote{\url{https://github.com/PeerOlafSiebers/ssc2026}}.

\section{Evaluation and Interpretation}
\label{sec:eval}

\subsection{Cross Human/LLM and LLM/LLM Response Comparison}
\label{sec:cross}

We compared the manually completed RAT-RS report of the Siebers and Aickelin \cite{siebers2011} paper against responses generated by four LLMs, including ChatGPT, Gemini, DeepSeek and Claude. A qualitative analysis of the 39 questions of the original RAT-RS for the human/LLM comparison and 44 questions of the extended RAT-RS for the LLM/LLM comparison revealed that the LLMs produced coherent, appropriately structured responses. Compared to human responses, these were semantically different. Consistent with Zeleke et al.\ \cite{zeleke2025}, human responses tended to be shorter, informal, analogy-rich, and conclusive, while LLM generated responses were longer, formal, policy-citing, and avoided drawing conclusions.

\begin{figure}[htbp]
  \centering
  \includegraphics[width=0.5\linewidth]{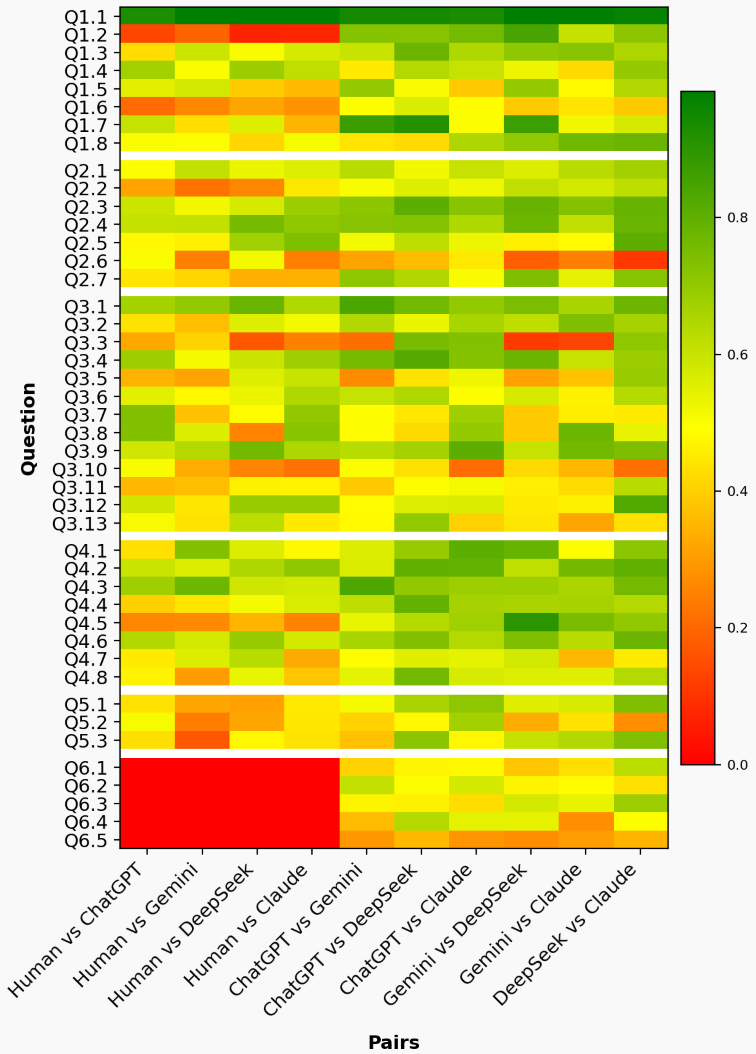}
  \caption{Semantic similarity heatmap for human/LLM and LLM/LLM responses.}
  \label{fig:heatmap1}
\end{figure}

The heatmap (Figure~\ref{fig:heatmap1}) provides information about the semantic similarity between the human/LLM and LLM/LLM responses. To generate it we calculated pairwise cosine similarity scores for all run-pair combinations across all questions. Cosine similarity (Manning et al. 2008) is a measure that captures the similarity of different vector profiles, which are - in our case - composed of the particular question item values. Note that Cosine similarity scores are used here only as a coarse indicator of topical overlap between texts, aiming at reflecting comparability of the general question profile, not as a measure of stylistic, structural, or rhetorical similarity (an aspect more relevant in classical qualitative content analysis). High values ($\geq 0.8$) indicate close semantic equivalence of response content, mid-range values ($0.6$--$0.8$) indicate partial overlap with divergent phrasing or reasoning paths, and low values ($< 0.6$) indicate semantic dispersion, not necessarily disagreement. All claims about comparative text properties beyond topical overlap would need to be grounded in separate analyses.

\paragraph{Human vs.\ LLM Evaluation.}
To measure semantic alignment between the human rater and each of the four language models, we compared pairwise cosine similarity scores for all four human-to-LLM pairs across 44 questions, with Q6.1-Q6.5 excluded from this subset as no human responses were recorded for those items. The average similarity score was 0.497 ($SD = 0.174$), indicating weak overall alignment and reflecting the genuine divergence between a human reviewer and LLM-generated responses. The tier distribution is strongly skewed: 77.3\% of questions fall in the Low tier, and only Q1.1 achieves High consistency. A Kruskal-Wallis test confirmed that differences across questions are significant ($H = 107.8$, $p < 0.0001$), meaning the question content, not the specific LLM being compared to the human, is the primary source of variability. Among the four Human-to-LLM pairs, column means range narrowly from 0.462 (Human vs.\ Gemini) to 0.511 (Human vs.\ DeepSeek), indicating that no single model is markedly closer to or further from the human rater than others. Gemini is the marginal outlier, though the gap is modest. Analysis by question type reveals the same gradient seen in the full dataset. The sole Factual question scored 0.969. Descriptive questions averaged 0.515 and Explanatory questions 0.420. The justification penalty in this subset, the drop associated with moving from descriptive to explanatory question types, is 0.095 points (Mann-Whitney $U = 221$, $p = 0.010$). This is substantially smaller than the equivalent figure in the LLM-to-LLM subset, not because the pattern is weaker, but because the overall score floor is lower: human and LLM responses diverge broadly across all open-ended question types, compressing their relative difference.

\paragraph{LLM vs.\ LLM Evaluation.}
To measure semantic alignment among the four language models independently of the human rater, we compared pairwise cosine similarity scores for all six LLM-to-LLM pairs across 44 questions. The average similarity score was 0.586 ($SD = 0.166$), meaningfully higher than the human-to-LLM mean of 0.497, and indicating moderate alignment among models that share broadly similar training conventions and response formats. The tier distribution is considerably less skewed than the human-inclusive subset: 52.3\% of questions fall in the Low tier, 45.5\% in the Medium tier, and one question achieves High consistency. A Kruskal-Wallis test confirmed significant between-question variation ($H = 157.6$, $p < 0.0001$). Among the six LLM pairs, column means range from 0.542 (Gemini vs. Claude) to 0.629 (DeepSeek vs. Claude), a spread of 0.087 points, wider than the human-to-LLM pair range of 0.049. DeepSeek and Claude show the strongest mutual alignment; Gemini is the least aligned with the other models across both its pairings. Analysis by question type replicates the full-dataset pattern with greater clarity. The sole Factual question scored 0.959. Descriptive questions averaged 0.666 and Explanatory questions 0.474. The justification penalty here is 0.192 points (Mann-Whitney $U = 288$, $p < 0.0001$), closely matching the 0.199-point penalty observed in the eight-run DeepSeek dataset. Evaluative questions averaged 0.459, comparable to Explanatory questions, and Binary questions averaged 0.528. The convergence of the justification penalty between the LLM-to-LLM subset and the intra-model repeated-run dataset is notable: when the human rater is removed from the comparison, the "what versus why" gradient becomes the dominant structural feature of the data in both, reinforcing the interpretation that this gradient is a property of the questions rather than of any particular respondent configuration.

\paragraph{Question Type Analysis.}
For the human/LLM comparison we can observe that for descriptive questions, LLM answers often matched or exceeded the human answer in detail, synthesising information from multiple paper sections. For explanatory questions, LLMs produced plausible but more variable answers, sometimes omitting nuances present in the human version. Notably, outright fabrications, such as claims made that were not present in the paper, were rare. Where they occurred, they were framed as inferences rather than confident statements, as the prompt instructed the model to indicate when information was missing or needed to be inferred. This confirms that LLMs are capable drafters but require human verification for interpretive questions. 

An example for the human/LLM comparison, the worst performing question is Q1.2. Here, the LLM answer clearly outperformed the one given by the human, even providing text quotations to justify the choice. Taking the Human-DeepSeek comparison as an example: Q1.2: What is the purpose of the model? (prediction, explanation, description, theoretical exploration, illustration, analogy, social interaction, or other (please specify). Human @ Prediction. DeepSeek @ Purpose: explanation and description. "To investigate the impact of people management practices on retail productivity" (abstract) and specifically "to investigate the impact that incorporating proactive staff behaviour … has on simulation output accuracy" (p.2). Also to improve model validity. As the Siebers and Aickelin (2011) paper is a methodology paper, DeepSeek's response provided the correct response. However, we can also observe cases where LLMs provide unnecessary complications. The semantic analysis in Q3.8 is one example. Here, human and LLM express the same content, but use different wording, with DeepSeek adding some conditional logic to the response. Q3.8: If you did not use it, why not? Human @ NOT APPLICABLE - Data was used when it existed. DeepSeek @ Conditional on Q3.7 = no; Q3.7 is not applicable, so this question is not applicable. Such cases might be alleviated with further fine-tuning of the extraction prompt. Those are just two examples that illustrate potential benefits but also challenges in using LLM-supported information extraction, motivating the importance of human oversight. Further illustrative examples showcasing the divergence in human and LLM response are provided in Section 4.3.

\subsection{Internal Consistency across LLM Responses}
\label{sec:intern}

To measure internal consistency, we ran the same RAT-RS prompt eight times on the same paper using DeepSeek and calculated pairwise cosine similarity scores for all 28 run-pair combinations across 44 questions. Each run was performed with a new context window to ensure the independence of the model runs. The heatmap (Figure~\ref{fig:heatmap1}) illustrates the degree of consistency achieved across repeated runs of the same model.

\begin{figure}[htbp]
  \centering
  \includegraphics[width=\linewidth]{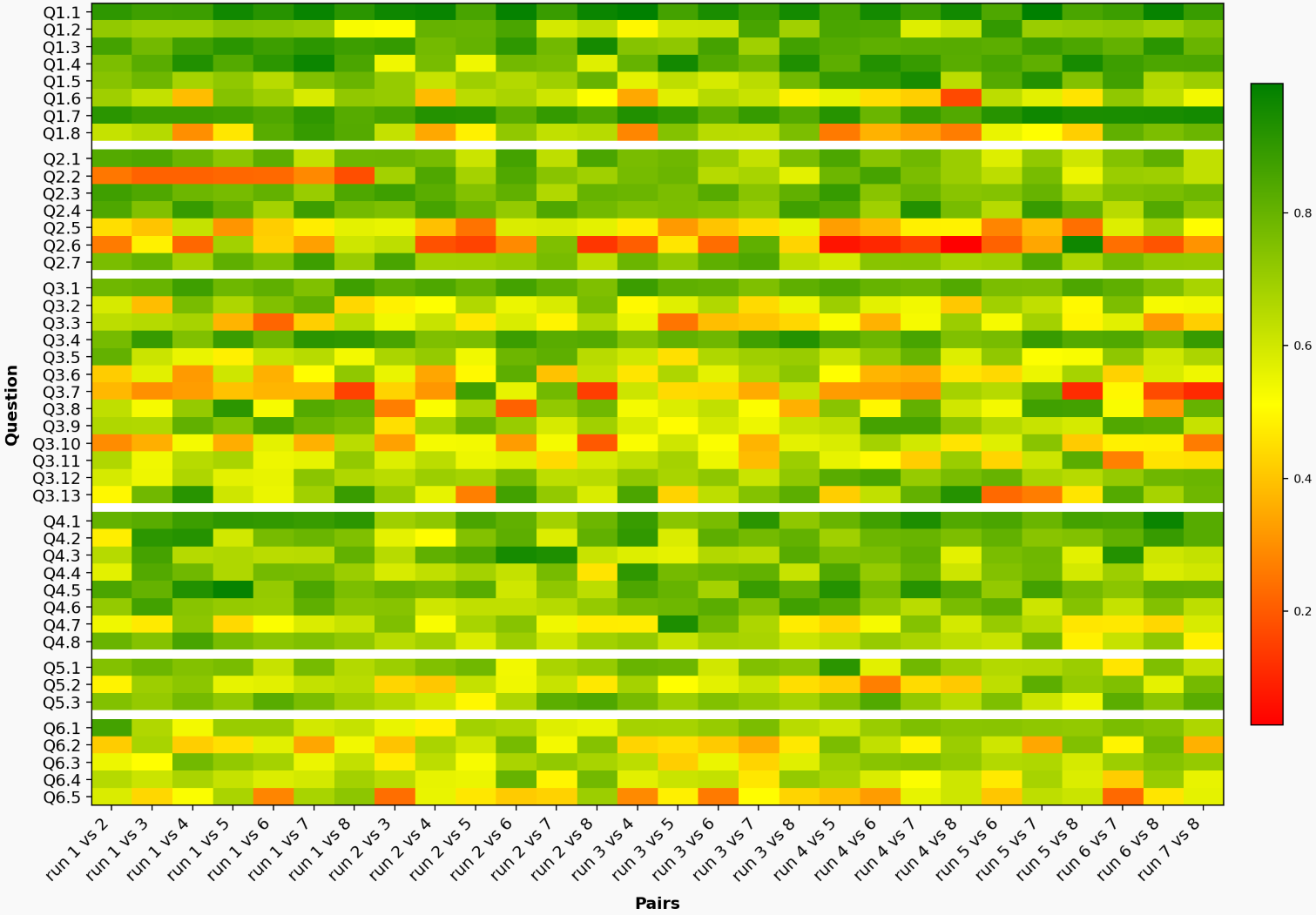}
  \caption{Consistency heatmap across repeated runs of the same model.}
  \label{fig:heatmap2}
\end{figure}

The average similarity score was 0.664 (SD = 0.175), indicating moderate overall consistency. However, this general figure hides a clear pattern. The Kruskal-Wallis test confirmed that the differences in consistency across questions are significant (H = 670.2, p < 0.0001): the model's reproducibility hence depends on the question type, not on the specific run. Among the 28 run pairs, the average similarity ranged from 0.631 to 0.700, indicating no individual run was a major outlier; variation from random sampling was well managed. The question itself is the source of inconsistency. Analysis of question types explains this trend. Descriptive questions, which ask what data was used and what sources were consulted, demonstrated a mean consistency of 0.750. Factual questions, such as the publication reference, scored 0.921. Explanatory questions, which require justifications, scored only 0.551. Evaluative questions, seeking assessments of reporting quality, scored 0.586. The difference between 'what' and 'why' drives reproducibility in this dataset. We refer to the 0.199-point mean drop linked to justification requirements as the justification penalty. (Mann-Whitney U = 294, p < 0.0001). This finding makes sense conceptually. Descriptive questions anchor responses to specific, identifiable sections of the paper: the answer is grounded in the text. Explanatory questions require the model to think about cause and effect, evaluate competing factors, and create a narrative. These cognitive tasks vary more across runs because the answer space is open-ended. This is not a flaw specific to LLMs; it mirrors how human reviewers differ on interpretive versus factual items in qualitative coding.

\subsection{Example Responses comparing Human and LLMs}
\label{sec:exam}

Augmenting this largely quantitative assessment with a qualitative view, we provide two random chosen examples to showcase typical patterns observable in the extraction foci of human and LLMs, respectively. To appreciate the nuance associated with the particular type and phrasing, we organise the examples by question type.

\paragraph{Descriptive example.}
Q3.2 Are these data elements implemented with the help of qualitative or quantitative data or further models? Response: Human @ Mix of quantitative (service time, arrival rates, probabilities), qualitative (states + transitions, proactive strategy, archetypes), and model-based (queuing behaviour, network) data. ChatGPT @ They used both quantitative and qualitative evidence. Quantitative data fed distributions and parameters, while qualitative observation and interviews supported state charts and rule design. Gemini @ Data elements are implemented using both quantitative data (rotas, company reports) and qualitative insights from observations and interviews (p.5, §3; p.6, §3). DeepSeek @ Quantitative data: frequency distributions (triangular), probability distributions, company transaction numbers. Qualitative data: customer type definitions from observations. No further models used. Claude @ Both. Quantitative: company transaction data, rota data, numerically-coded observation timings, survey results. Qualitative: semi-structured interviews and participant observation informing behavioural rules and agent state designs. No further sub-models are used as inputs. (p.5, §6; p.10). The descriptive example, with focus on offering descriptive account of data use in the model, highlights the general similarity in content, yet variation in expressiveness, with the human being factually oriented, and LLMs often more elaborate (albeit with pointy responses at time (see e.g. Claude's "Both."). Noteworthy is the partially stronger anchoring on concrete evidence based on paragraph and page references.

\paragraph{Explanatory example.}
Q3.3 Explain how data affected the way you implemented each model element and why. (i.e. explain your choice of data elements). Response: Human @ Data elements chosen based on: required for modelling service provided, experimental factors for proactive behaviour, required to replicate real system operations and behaviours, available and suitable for validation. ChatGPT @ Because the empirical sample was small and numerical, they used triangular distributions for durations; probabilities came partly from company data and partly from informed guesses; survey results set satisfaction weights; managers defined customer mixes; observations shaped proactivity rules. Gemini @ Discrepancies between initial model results and real performance data led authors to implement proactive "role swapping" mechanisms to better represent true operational procedures (p.2, §2). DeepSeek @ Chose triangular distributions because "we have only a relatively small sample of empirical data and a triangular distribution is commonly used as a first approximation" (p.10). Probabilities from company data (e.g., conversion rates) and informed guesses (e.g., patience). Claude @ Numerical data were implemented as statistical distributions 'as statistical distributions are the best format to represent the data we have gathered … due to their numerical nature' (p.10). Where only small samples existed, triangular distributions were used. Where numerical data were absent, probability values were set from expert estimates and company data combined. (p.10). The response style for explanatory questions more or less consistently appears relatively verbose, given the models' focus on referring to sentences or statements that elicit explanatory insights. At the same time, the focus on evidence-based reference is visible. LLMs produce more concrete, evidence-linked justifications, yet their variability suggests human oversight remains critical for interpretive accuracy.

Overall, this study, covering both quantitative and qualitative aspects, alongside the differentiated results based on different question types allow us to draw specific insights highlighted in the following section.

\subsection{Summary and Lessons Learned}
\label{sec:summary}

The systematic comparisons in Section~\ref{sec:cross} shows that responses for 43 out of 44 questions fall into the low similarity tier when human responses are compared with those generated by an LLM. As mentioned in those sections, this outcome is largely attributable to systematic stylistic divergence between human and machine-generated text, not to real differences in the information extracted. This stylistic difference lowers cosine similarity scores regardless of the actual meaning and should therefore not be seen as proof of informational disagreement. Importantly, the inter- and intra-LLM analyses in Sections~\ref{sec:cross} and~\ref{sec:intern} offer a different view. Moderate-to-high consistency across four different LLMs, and even higher consistency with repeated runs of the same LLM, suggest that the information being extracted is broadly convergent. For factual and descriptive questions in particular, the evidence indicates that divergence operates primarily at the level of linguistic formulation rather than informational content. This distinction has clear practical implications: it supports using LLM-generated drafts as a reliable starting point for supervised extraction workflows, as long as human review focuses on interpretive accuracy rather than just surface-level wording.

\section{Conditions and Caveats: Where the Model Holds}
\label{sec:caveats}

We do not believe that LLMs can completely replace human judgement in RAT-RS completion. Instead, we think they can reliably handle central aspects of the extraction work if three conditions are satisfied.

\paragraph{First, humans must stay involved.}
The consistency analysis gives an evidence-based way to guide review: descriptive questions, such as the ones in the Calibration and Experiments block (mean consistency 0.731) and the Administrative block (0.921) of the RAT-RS, can be considered high-confidence outputs that require minimal review. In contrast, explanatory and evaluative questions, particularly those in the Data Existence and Collection block (mean 0.550) and the Reflective Appraisal block (0.586), should be flagged for thorough human review, ideally with multiple runs compared before accepting a final answer. The question-type taxonomy acts as a triage system, focusing human attention where it is needed most. The two lowest-performing items, Q14 (mean similarity 0.355, coefficient of variation 0.677) and Q22 (mean 0.423, CV 0.491), consist of explanatory or binary-conditional questions that require justification for decisions about absence or non-use. These questions often have vague or incomplete data reporting in ABM publications. The model's inconsistency might reflect real ambiguity in the source text rather than a failure of the model. This is helpful diagnostically; high variability across runs signals to the reviewer that the underlying information is truly contested or missing.

\paragraph{Second, LLM use must be disclosed.}
A supervised extraction workflow changes how we see the RAT-RS report. It is no longer just a first-person account from the researcher; it becomes a validated extraction produced with AI help. This distinction should be clear in the document. We suggest that reports created under supervised extraction include standard information, such as the model used, the prompt version, the number of runs, and the nature of the human review involved. 

\paragraph{Third, the prompt must be versioned and shared.}
The quality of the extraction relies heavily on prompt design. Improving the prompt should be a community effort, not just an individual task. A prompt developed through diverse papers, modelling traditions, and data types will outperform any single-authored version. Consistency and comparability is maintained via explicit conventions for prompt versioning. This requires a shared repository, a versioning system, and a community culture of contributing validation cases.

\section{Discussion and Future Work}
\label{sec:discussion}

This study provides a feasibility assessment of LLM-assisted supervised extraction for RAT-RS-based ABM documentation. The results suggest that LLMs can reliably support structured data-use reporting under human supervision, particularly for descriptive and factual question types, while performance is less stable for explanatory and evaluative items. Our analysis highlights systematic variation in consistency across question types, supports the use of multi-run and cross-model comparisons, and indicates that LLMs are best positioned as drafting and extraction assistants rather than autonomous reporters. The study also identifies key methodological constraints related to generalisability, reproducibility, and evaluation metrics, and outlines pathways for improving robustness through automation, prompt engineering, and structured workflows.

Given the nature and motivation of this paper, we want to be explicit in acknowledging practical, methodological and technical limitations. Firstly, this study is based on a single-paper feasibility design and does not claim statistical generalisability across the broader ABM literature. The selected case (Siebers and Aickelin, 2011 \cite{siebers2011}) was included because it provides an existing RAT-RS benchmark for comparison, not because it is representative of typical ABM reporting practices. Consequently, results may vary substantially for studies with richer empirical grounding, more complex data integration, or limited methodological transparency, which may affect both absolute consistency levels and observed differences across question types. Associated with this, we focus on the RAT-RS standard as an example of a community-based standard, but note that the principal approach, with appropriate modification to the extraction prompt, is applicable to other protocols and standards, such as ODD; an extension we see as subject to future work and, more desirably, community effort (see Section~\ref{sec:action}). A further, methodological limitation concerns the use of cosine similarity as the primary measure of response alignment. This metric captures only topical overlap and should not be interpreted as a measure of correctness, stylistic similarity, or argumentative equivalence. It therefore provides a coarse proxy for convergence in extracted content rather than a full validation of semantic equivalence. Finally, reproducibility is constrained by the use of commercial LLMs, which are non-deterministic and subject to version changes over time. This reinforces the need for multi-run sampling, explicit prompt versioning, and systematic metadata reporting (e.g.\ model version, sampling settings, and run count reporting) as core components of any supervised extraction workflow. In addition, it encourages community efforts to self-host or provide community-hosted variants to ensure stronger control about the used variants and versions.

Let's return to the bigger picture. This paper assumes a constructive position towards the adoption of LLMs. It is not about whether, but where LLMs can support the research process. This paper focuses on documentation practices and makes an associated feasibility, methodological and diagnostic claim, and provides initial evidence and pathways of application, highlights limitations, as well as opportunities transcending beyond the specific standard employed for the study. This includes the general feasibility of LLM-assisted generation or augmentation of data usage protocols under human guidance. We do not claim statistically robustness for general cases (subject to future work) but show the circumstances under which it can be applied. Associated with this, we introduce starting points for methodologically rigorous application of such approach based on intra- and inter-model assessments, as well as procedural documentation practices (e.g.\ prompt, meta-data, automation). Finally, this initial study reveal diagnostic opportunities regarding the ability of LLMs to support extraction based on particular question types, offering a nuanced understanding of limitations of applicability, while helping to develop heuristics that can inform the use of LLM-assisted information extraction for other ABM-related documentation standards that likewise rely on distinctive question types. All these insights -- the adoption of documentation standards, the opportunities but also challenges associated with the presence of LLMs -- encourage us to synthesise this into a tangible call to action addressed at the broader community:

\section{Call to Action}
\label{sec:action}

We encourage the SSC community to take three specific steps to support the exploration of LLM adoption for documentation purposes.
\textbf{Adopt the prompt template and toolbox} to support the systematic systematic extraction of information of data use. The extraction prompt, visual analysis tool, and support scripts are available on GitHub and open to further refinement. Researchers should apply the supervised extraction workflow to their own ABM papers, both published and in preparation, and share their experiences. Even a few cases of adoption will provide evidence about where the workflow works well, where it does not, and what prompt improvements are most beneficial. 
\textbf{Contribute validation cases}, as the feasibility study is based on a single paper with a known accurate result. To generalise its findings, we need a larger body of work. We invite community members to run the prompt on their papers and submit the outputs, along with their assessments of accuracy, to a shared repository. This crowdsourced approach to validation, similar to initiatives in systematic review automation, is the quickest path to a strong, field-validated prompt.
\textbf{Update standards and protocols}, such as ODD and the RAT-RS to officially include supervised extraction. The current RAT-RS specifications, for instance, do not cover LLM-assisted generation. We suggest that authors of standards and the broader community discuss a formal change that (a) defines acknowledgement protocols for AI-assisted reports, (b) identifies which question types and blocks can be extracted confidently using automation and which need mandatory human review, and (c) considers whether rethinking questions, such as breaking apart explanatory items, supporting binary-conditional items, and adding rubrics to evaluative items, could improve consistency for both human and LLM responses. 

Supervised extraction models do not resolve every issue in ABM model documentation or data reporting. They cannot ensure that sparse documentation turns into rich documentation or that incomplete papers suddenly provide complete answers. However, it changes the practical threshold for feasibility and helps researchers, whether novices or experts, to recognise gaps in their documentation they might otherwise overlook. Now, producing a documentation report, such as RAT-RS or ODD report, is achievable for a researcher with strong time constraints, but access to a general-purpose LLM and a community-developed practice that supports rigorous use. That is not a small improvement. It is a step marking the difference between standards that are well-discussed, yet aspirational in practice and ones that are actually attainable.

\bigskip
\noindent\textbf{Generative AI Disclosure.} Generative AI tools were used
in the preparation of this manuscript to support copy editing and the
development of statistical analysis scripts.

\bigskip
\noindent\textbf{Supplementary Material.} The extraction prompt toolbox for
RAT-RS (all flavours), the generated data used for the semantic analysis,
and the Python script used for generating statistics are available on
GitHub (\url{https://github.com/PeerOlafSiebers/ssc2026}).


\end{document}